\documentclass[prl,twocolumn,superscriptaddress]{revtex4}

\def\be{\begin{equation}}
\def\ene{\end{equation}}

\begin{document}

\title{The dynamical foundation of fractal stream chemistry: The origin
of extremely long retention times}

\author{H. Scher}
\email{harvey.scher@weizmann.ac.il}
\affiliation{Department of Environmental Sciences and Energy Research,
Weizmann Institute of Science, Rehovot, 76100, Israel}
\author{G. Margolin}
\email{g.margolin@weizmann.ac.il}
\affiliation{Department of Environmental Sciences and Energy Research,
Weizmann Institute of Science, Rehovot, 76100, Israel}
\author{R. Metzler}
\email{metz@nordita.dk}
\affiliation{Department of Physics, Massachusetts Institute of Technology, 77
Massachusetts Avenue, Cambridge, MA 02139}
\affiliation{NORDITA, Blegdamsvej 17, DK-2100 Copenhagen {\O}, Denmark}
\author{J. Klafter}
\email{klafter@post.tau.ac.il}
\affiliation{School of Chemistry, Tel Aviv University, 69978 Tel Aviv, Israel}
\author{B. Berkowitz}
\email{brian.berkowitz@weizmann.ac.il}
\affiliation{Department of Environmental Sciences and Energy Research,
Weizmann Institute of Science, Rehovot, 76100, Israel}

\begin{abstract}
We present a physical model to explain the behavior of long-term, time series
measurements of chloride, a natural passive tracer, in rainfall and runoff in
catchments [{\it Kirchner et al.}, {\it Nature, 403}(524), 2000]. A spectral
analysis of the data shows the chloride concentrations in rainfall to have a
white noise spectrum, while in streamflow, the spectrum exhibits a
fractal $1/f$ scaling. The empirically derived distribution of tracer 
travel times
$h(t)$ follows a power-law, indicating low-level contaminant delivery to
streams for a very long time. Our transport model is based on a continuous time
random  walk (CTRW) with an event time distribution governed by $\psi(t)\sim
A_{\beta}t^{-1-\beta}$. The CTRW using this power-law $\psi(t)$ (with 
$0<\beta<1$)
is  interchangeable with the time-fractional advection-dispersion 
equation (FADE) and
has accounted for the universal phenomenon of anomalous transport in 
a broad range
of disordered and complex systems. In the current application, the 
events can be
realized as transit times on portions of the catchment network.
The  travel time distribution is the first passage time distribution 
$F(t; l)$ at a
distance $l$ from a pulse input (at $t=0$) at the origin. We show 
that the empirical
$h(t)$ is the catchment areal composite of $F(t; l)$ and that the fractal 
$1/f$ spectral
response found in many catchments is an example of the larger class of
transport phenomena cited above. The physical basis of $\psi(t)$, which
determines $F(t; l)$, is the origin of the extremely long chemical 
retention times in
catchments.
\end{abstract}

\maketitle

\section{Introduction}

The distribution of travel times required for chemical substances to travel
through a catchment is an important hydraulic quantity, determining the
retention of pollutants until they are eventually released to a stream or lake.
The same distribution controls the transport of chemicals in subsurface
hydrological systems. The nature of this distribution determines the
expected ecological impact of contaminants.

In a long-term catchment study conducted at Plynlimon, Wales,
{\it Kirchner et al.} [2000] (hereafter denoted KFN) compare the (input) time
series of the concentration
$c_R(t)$ of
chloride tracer in the rainfall to the (output) time series of the
concentration
$c_s(t)$ into the Hafren stream. They relate the concentrations through the
convolution integral
\begin{equation}
\label{conv}
c_s(t)={\textstyle\int_0^\infty}h(t')c_R(t-t')dt'
\end{equation}
where the effective travel time distribution $h(t)$ governs the lag
time between
injection of the tracer through rainfall and outflow to the stream.

KFN observe that the streamflow (volume of water) responds
rapidly to storm rainfall  inputs while the
$c_s(t)$ response is highly damped (i.e., low-level contaminant
delivery to streams
for a very long time). The timescales of catchment hydrologic and
chemical response
are resolved using spectral methods \cite{duffy} (in which the 
input/output response
can be compared at each (time) wavelength). The spectral power of the 
water fluxes of
rainfall and streamflow coincide while the spectra of $c_s(t)$ show strong
attenuation on all wavelengths less than 5-10 years. Further, the
$c_R(t)$ spectra
scale as white noise in sharp contrast to the fractal $1/f$ scaling of the
$c_s(t)$ spectra. Using these results in the Fourier transform of the
convolution in (\ref{conv}), with $f$ denoting frequency, and
$C_s$, $H$ and $C_R$ representing the respective Fourier transforms,
\begin{equation}
\label{amp}
C_s(f)=H(f)C_R(f)
\end{equation}
and the relation of the power spectra is
\begin{equation}
\label{power}
|C_s(f)|^2=|H(f)|^2|C_R(f)|^2
\end{equation}
Since $|C_R(f)|^2$ is nearly a white noise spectrum (a constant) one has
$|H(f)|^2 \propto |C_s(f)|^2$. From the measured power-law scaling of
$|C_s(f)|^2
\propto f^{-0.97}$, KFN conclude that
\begin{equation}
\label{htail}
h(t) \sim t^{-m}
\end{equation}
(from $H(f)\sim f^{-(1-m)}$), where $m \approx 0.5$.

In order to ensure that $h(t)$ is integrable and possesses a finite
average travel
time, KFN arbitrarily chose the
gamma distribution
\begin{equation}
\label{tddk}
h(t)=\frac{t^{\gamma-1}}{\varrho^{\gamma}\Gamma(\gamma)}
e^{-t/\varrho}
\end{equation}
with $\gamma=0.48$ and $\varrho=1.9$ years, with a characteristic travel time
$T_h \equiv\int_0^{\infty}h(t)t dt=\gamma\varrho$.
Due to the relatively large value of $\varrho$, essentially the entire data
correspond to the power-law $h(t)\sim t^{\gamma-1}$ with the
power spectrum $|C_s(f)|^2 \propto|H(f)|^2\sim f^{-2\gamma}$.
KFN refer to similar scaling between $f^{-0.7}$ and $f^{-1.2}$ found in
Scandinavian and North American field sites, which indicates a certain
ubiquity to fractal, scale-free forms (which is the issue we address in this
letter).
\section{Transport models}

The first step to understanding transport in the catchment is the
clarification of
the meaning of (\ref{conv}). The $h(t)$ is the {\it effective}
response to a pulse
of rain falling on the entire area of the catchment. Every point of
this area is a
source of chloride and the stream is a line sink for the chloride. We
simplify the
area to be a rectangle of width $\lambda$ about this stream sink (with periodic
boundary conditions on the sides perpendicular to the
stream\cite{ctrw}). The sink
or absorbing boundary ensures the correct ``counting rate" of chloride at the
arriving point, i.e., the  first passage time distribution $F(t;l)$ is the
appropriate travel time distribution from a pulse source at point
$l$. In terms of
this {\it intrinsic} distribution,
\begin{equation}
\label{eq:gen}
c_{s}(t;l_s) = \int_0^\infty \sum_{l\in \Omega}F(t';l-l_s) c_{R}
(t-t';l)dt'
\end{equation}
where $c_{s}(t;l_s)$ is the chloride concentration at the stream
position $l_s$ at
time $t$, $\Omega$ is the size of the catchment and $c_{R}(t;l)$ the
rain-input at a position $l$ in $\Omega$. We can consider the
sampling positions
$\{l_s \}$ of
$c_{s}$ to be a small region compared to $\Omega$ and hence
$c_{s}(t;l_s) \approx
c_{s}(t)$.  To do the $l$-sum we assume that the
rainfall is distributed uniformly in $\Omega$, $c_{R}(t;l) \approx
c_{R}(t)$. Hence,
we recover (\ref{conv}) now with
\begin{equation}
\label{eff}
h(t) \equiv \sum_{l\in \Omega}F(t;l)
\end{equation}
The distribution $F(t;l)$ must be integrable in time and thus so $h(t)$ for a
finite $\Omega$. The basis of comparison for various transport
approaches is the
computation in (\ref{eff}).

\subsection{Advection-dispersion equation (ADE)}

The textbook approach to transport of passive tracers in both surface
and subsurface
systems usually focuses on the advection-dispersion equation (ADE)
\begin{equation}
\label{dae}
\partial {\cal W}/\partial t+v\partial {\cal W}/\partial x=D\partial^2 {\cal
W}/\partial x^2
\end{equation}
e.g., for 1d with constant $v$ the average fluid velocity and $D$
the dispersion coefficient. This equation governs the temporal evolution
of the probability density function (pdf) ${\cal W}(x,t)$ of finding a
tracer particle,
which undergoes dispersive motion, at a certain
position $x$ at time
$t$ after release at $t=0$. The pdf ${\cal W}$ is the normalized
concentration profile.

The travel time distribution belonging to (\ref{dae}) can be obtained
by the Laplace
Transform ($LT$) technique. This yields the $LT$
$\tilde{{\cal F}}(u;x)\equiv\int_0^{\infty}{\cal F}(t;x)e^{-ut}dt$ of the first
passage time from
$x$ to the absorbing boundary at $x=0$, $\tilde{{\cal F}}(u;x)=
\exp\left(x\left[v-\sqrt{4Du+v^2}\right]/(2D)\right)$. $LT$ inversion
delivers the travel time distribution
\be
\label{gauss}
{\cal F}(t;x) = \frac{x}{\sqrt{4\pi
D}t^{\frac{3}{2}}}\exp\left(\frac{-(vt-x)^2}{4Dt}\right)
\ene
which when inserted for $F(t;x)$ in (\ref{eff}) yields
\begin{eqnarray}
\label{int}
h(t) \! \! \! \! &\propto& \! \! \! \! \int_0^\lambda {\cal F}(t;x)dx =\frac
{v}{2}\bigl[{\rm erf}(z) +{\rm erf}\left(\frac {P_e}{4z}-z \right) 
\nonumber \\&+&
\! \! \! \! \frac {\exp(-z^2)-\exp\left(-(\frac
{P_e}{4z}-z)^{2}\right)}{\sqrt{\pi}z}\bigr]
\end{eqnarray}
where the Peclet number $P_e\equiv \lambda v/D$, 
$z \equiv \frac {1}{2}\sqrt{v^{2}t/D}$ and
erf$(z)$ is the error function (the same result (\ref{int}) was also 
derived in {\it
Kirchner et al.}, 2001). For
$t<0.1 D/v^2$, $h(t)\sim t^{-\frac{1}{2}}$ which is similar to 
(\ref{htail}) for $m
\approx 0.5$. [This upper limit on the time range holds for 
$P_e\ge1$; for $P_e<1$ the limit is smaller.] The range is estimated 
by assuming $D$
to be the macrodispersion, $D=v\alpha$, where $\alpha$ is the dispersivity.
Assuming $v\approx 10^2$ m/yr and on the km scale $\alpha \sim (1-10^2)$ m,
$t<0.1\alpha/v \sim (10^{-3}-10^{-1})$ yr. One needs $v\approx 10$ 
m/yr and $\alpha
\sim 10^2$ m in (\ref{int}), for $h(t) \sim t^{-\frac{1}{2}}$ to 
partially overlap
the time range of observation of this dependence \cite{kirchner}. 
Moreover, this
behavior is only the same as (\ref{htail}) in the special case where $m$ (or
$\gamma$) = $\frac {1}{2}$. The catchment data (spectral power $\sim 
f^{-2\gamma}$)
quoted above  cover a range of
$0.7\leq 2\gamma\leq 1.2$.

For large $t$, $h(t)$ in (\ref{int}) exhibits an exponential decrease
$\sim t^{-\frac{3}{2}}$ $\exp\left(-v^2(4D)^{-1}t\right)$ which is faster than
(\ref{tddk}). The decrease in $h(t)$ at large $t$ ($t>>\lambda/v$) is
due to the finite size of the catchment and not an arbitrary limiting time. The
relation between the operative time range of (\ref{htail}) and
$\lambda$ will be
determined in the following.

\subsection{Continuous-time random walk and time-fractional-ADE}

As shown above the standard treatment of the transport using the ADE
yields a marginal accounting of the data. A transport model must determine
a $F(t;l)$ that when inserted in (\ref{eff}) gives rise to 
(\ref{htail}) for over
three decades of time \cite{kirchner}. As we will show this is best 
expedited with
a non-Gaussian form of $F(t;l)$. A particularly dispersive form of $F(t;l)$ is
associated with anomalous transport which has been observed in a wide range of
disordered and complex systems: electron hopping/multiple-trapping in amorphous
semiconductors \cite{ctrw}, particle migration in fracture networks 
\cite{till},
contaminant transport in heterogeneous porous media \cite{bescher},
anomalous diffusion\cite{report}, and Hamiltonian chaos\cite{pt}. There
are key common features in these anomalous transport phenomena, e.g., 
non-Gaussian
propagation of an initial pulse of particles with a mean position $\ell(t)$ and
standard deviation $\sigma(t)$ exhibiting the same sublinear 
dependence on $t$ (in
the  presence of a bias or head). A continuous time random walk 
(CTRW) transport
process has successfully accounted for these unusual basic features. The
anomalous behavior is an outcome of a CTRW governed by a long-tail 
distribution of
the individual transition times or event times (which have to be 
defined in each
physical context, e.g., trap release),
\begin{equation}
\label{psi}
\psi(t)\sim A_{\beta}t^{-1-\beta}
\end{equation}
where $A_{\beta}$ is a constant.
Over the observation time range in which (\ref{psi}) obtains we have for the
propagating plume $P(l,t)$
\begin{equation}
\ell (t) \sim t^\beta, \qquad\sigma(t) \sim t^\beta\qquad0<\beta<1
\end{equation}
The highly dispersive nature of this $P(l,t)$ can be discerned by the ratio
$\sigma(t)/\ell (t)\sim$ constant, instead of the familiar $1/\sqrt{t}$ for a
Gaussian plume. The latter is produced for $\beta>2$ (while for
$1<\beta<2$ one has an intermediate $t$-dependence).

The CTRW accounts naturally for the cumulative effects of a sequence of
transitions which constitutes the particle transport. The formalism
of the CTRW has
been well documented in the literature [{\it Berkowitz and Scher}, 1998;
{\it Metzler and Klafter}, 2000]. For brevity we
show the key equation for our purpose here in Laplace space (for
three dimensions)
\begin{equation}
\label{1stP}
\tilde{F}({\bf s},u)=\tilde{P}({\bf s},u)/\tilde{P}(0,u) \qquad
\qquad {\bf s} \ne 0
\end{equation}
where the particle starts at the origin at $t=0^+$, $\tilde{P}({\bf
s},u)$ is the
$LT$ of $P({\bf s},t)$, the probability density to find the particle
on ${\bf s}$ at
time $t$ (plume), and $\tilde{F}({\bf s},u)$ is the $LT$ of $F({\bf s},t)$, the
probability per time for the particle to first arrive at ${\bf s}$ at time $t$.

Explicit evaluations for $\tilde{F}({\bf s},u)$ have been determined
for symmetric
situations (e.g., plane (line) source to plane (line) sink) and with the use of
(\ref{psi}) \cite{ctrw}. The symmetric cases allow a reduction of the problem
to a one-dimensional form of $F$ which we write as $F(t; x)$ and
\begin{equation}
\label{ltF}
\tilde{F}(u;x)=\exp(-xu^{\beta}/\bar{l}) \qquad(0<\beta<1)
\end{equation}
($u$ dimensionless). The inverse $LT$ of $\tilde{F}(u;x)$ has been
expressed in terms
of a class of Fox $H$-functions \cite{report}, but it is more expedient to work
directly with (\ref{ltF}) in the evaluation of $h(t)$, which will be
considered in
the next section.

Mathematically, the CTRW (in the special case using (\ref{psi}) and
$(0<\beta<1)$)
is interchangeable with the time-fractional advection-dispersion
equation (FADE)
\cite{report}
\begin{equation}
\label{fda}
\frac{\partial W}{\partial t}=\, _0D_t^{1-\beta}\left(-v_{\beta}
\frac{\partial}{\partial x}+d_{\beta}\frac{\partial^2}{\partial x^2}
\right)W(x,t)
\end{equation}
where the Riemann-Liouville operator is defined in terms of the convolution
\cite{report},
\begin{equation}
\label{rl}
_0D_t^{1-\beta}W(x,t)=\frac{1}{\Gamma(\beta)}\frac{d}{dt}\int_0^tdt'
\frac{W(x,t')}{(t-t')^{1-\beta}},\, 0<\beta<1.
\end{equation}
The form of FADE (\ref{fda}) is the natural generalization of the ADE
(\ref{dae}) for power-law forms of $\psi$. The solutions of FADE
reproduce those of
the CTRW in this special case, and in the spatial continuum limit they
form a natural
bridge between the CTRW and the ADE \cite{tipm}.

\section{Long Retention Times}
We proceed to insert (\ref{ltF}) into the $LT$ of (\ref{eff}) to obtain
\begin{equation}
\label{eq:lth}
\tilde{h}(u) \propto \int_0^\lambda
\exp(-lu^{\beta}/\bar{l})dl=\bar{l}u^{-\beta}(1-\exp(-\lambda
u^{\beta}/\bar{l}))
\end{equation}
where $\bar{l}$ is the mean step distance. The expression for $\tilde{h}(u)$ in
(\ref{eq:lth}) has been  thoroughly studied in
another context \cite{ctrw}; the main features are
\begin{equation}
h(\tau)\sim\left\{\begin{array}{ll}
\tau^{\beta-1}, & \tau<\tau^*\\
\tau^{-\beta-1}, & \tau>\tau^*
\end{array}\right.
\label{scaling}
\end{equation}
($\tau$ dimensionless time). The exponent for $\tau>\tau^*$ ensures that
$h(\tau)$ is
integrable. On a log-log plot $h(\tau)$ is two constant slopes, $\beta-1$,
$-1-\beta$, with a turnover range between them.  The center time of this range,
$\tau^*$, can be estimated as the
time for the argument of the exponent in (\ref{eq:lth}) to be $\sim
O(1)$ (using $u \sim 1/\tau$)
\begin{equation}
\label{tau}
\tau^{*} \sim (\frac {1-\beta}{\beta})^{\frac {1-\beta}{\beta}}(\beta
\lambda/\bar{l})^{\frac {1}{\beta}}
\end{equation}
(The $\beta$-factors derive from a more detailed analysis [{\it Scher
 and Montroll}, 1975, Appendix C].)
For $\beta =\frac{1}{2}$, in (\ref{eq:lth}) one can do the inverse
$LT$ and obtain $h(\tau) \sim (\pi
\tau)^{-\frac{1}{2}}(1-\exp(-\lambda^2 /4
\bar{l}^2 \tau))$, clearly showing (\ref{scaling})-(\ref{tau}).

The change from $\tau$ to $t$ is model dependent (to be discussed
below). We use $\tau = v_ot/\bar{l}\varepsilon$, where $v_o$ is a 
characteristic
velocity (of the velocity distribution) and $\varepsilon$ is a constant. For
$\beta = \frac{1}{2}$, $v_o \sim$ 100 m/yr, $\lambda \sim \frac{1}{2}$
km,  $\bar{l} \sim$ 30 m and $\varepsilon \sim \frac{1}{2}$, one has (from
(\ref{tau}) with $\tau^* = v_ot^*/\bar{l}\varepsilon$), $t^* \sim$ 10 
yr, which is a
reasonable time scale for the change from $h(t) \sim t^{\beta-1}$. Hence, our
scaling result (\ref{scaling}) agrees with the KFN data
(\ref{htail}) over the measurement time range ($>$ 3 decades for 
t$\le$ 10 yrs) with
$\beta=\gamma$ ($\equiv  1-m$). The $t$-``cutoff" for $h(t)$ in (\ref{scaling})
is algebraic, $t^{-\beta-1}$, not exponential as in (\ref{tddk})! 
This behavior of
$h(t)$ is  indicative of extremely long chemical retention times in 
catchments. The
turnover to the $t^{-\beta-1}$ dependence (in (\ref{scaling}))  is a 
prediction of
the CTRW theory and has not yet been observed.  A finite value of
$T_h$ depends on a change in the behavior of (\ref{psi}). For
$t
\gg t^*$, the tail of $\psi(t)$ becomes steeper and for $\beta>2$, the {\it
intrinsic} arrival time distribution evolves to ${\cal F}(t;x)$.

Independent tests of our transport model can be obtained with experiments
using a different tracer (d) and boundary conditions. For injection 
of the tracer
d at various catchment points $\{ {\bf x}
\}$, the concentration in the stream $c_{d}(t;{\bf l_s})$ is 
proportional to $F({\bf
x}-{\bf l_s},t)$,  determined for these boundary conditions. [This 
result answers
the question raised by {\it Stark and Stieglitz} [2000] about a site-specific
spill.] This is
also a method for obtaining a value of $\beta$, which determines the
shape of $F({\bf x},t)$.

There are many different physical mechanisms (cf. Refs. above) that
can give rise to the $\psi(t)$ in (\ref{psi}) that has generated our 
results for
$h(t)$. The common features are the representation of the transport 
process as a
series of transitions [{\it Berkowitz and Scher}, 1998, 2001] 
and the encounter within this
series of a sufficient number of transition-times that are much larger than the
median one. These relatively few long-time events, which can be due 
to release from
traps/stagnant-regions and/or passage through low velocity segments, can have a
dominant influence on the overall transport. The relative weighting
of these events is expressed by the exponent of the power-law in (\ref{psi}).

The hydrologic environments beneath the catchment subregions can 
contain a sufficient
number of these low velocity sections and stagnant dead-ends. The 
local slope (and
interconnection \cite{rinaldo}) of the catchment side-channels also affects the
velocity distribution. The subsurface of the catchment basin can be 
modeled as a
heterogeneous porous medium and/or a random fracture network 
\cite{bescher}. In the
latter the fragment-length distribution, $f(s)$, and the 
fragment-velocity histogram,
$\Phi({\bf \xi})$, are mapped onto the probability rate for a transit 
time t (through
a fragment) with a  displacement {\bf s},
\be
\psi({\bf s},t) \propto \Phi({\bf \xi})f(s)
\ene
where $|{\bf \xi}|=1/v$, $\hat{\bf \xi}=\hat{\bf v}$, $t=s\xi$ and
$\psi (t) \equiv \sum_s \psi ({\bf s},t)$. The anomalous transport 
observed in the
fracture network particle tracking simulations is due to the power-law tail of
$\Phi({\bf \xi}) \rightarrow \xi^{-1-\beta}$ at large $\xi$. This is 
the expected
behavior to be found in the catchment subsurface $\Phi_c({\bf \xi})$
distribution because this hydrologic environment is the same as ones modeled
successfully by the CTRW theory  [{\it Berkowitz and Scher}, 1998,
2001]. The experiments outlined above can test these properties.

\section{Conclusion}

In this letter we introduce the phenomenon of anomalous transport
into an analysis of the comparison of the power spectra of the time series of
chloride in rainfall and in (stream) runoff. An {\it effective} distribution of
travel times $h(t)$ is derived (by KFN from the data) which we show 
is composed of
the summation from all sites of the source distribution $F({\bf s},t)$ in the
catchment. The  latter, in turn, is composed of a sequence of 
transitions governed by
a power-law (tail) distribution (\ref{psi}). The results agree naturally
with the data reported from a number of catchment studies, in the sense of a
power-law scaling of $h(t)$ over decades of $t$ in the observational time range
(using reasonable parameters). A consequence of our transport model is a
power-law "cutoff" of $h(t)$ predicting extremely long chemical 
retention times with
subsequent impact on the long-term effects of contaminants on these ecosystems.

\acknowledgments
We thank James W. Kirchner and Colin Stark for a critical reading
of the manuscript. RM acknowledges the DFG and the European Commission
within the Emmy Noether and
SISITOMAS programmes, and BB acknowledges the European
Commission (Contract EVK1-CT-2000-00062), for financial support.

\end{document}